\documentclass[twocolumn,pre,superscriptaddress,10pt,aps,floatfix]{revtex4-2}
\usepackage[english]{babel}
\usepackage{graphicx}
\usepackage{amsmath}
\usepackage[colorlinks,linkcolor=blue,citecolor=blue,urlcolor=blue]{hyperref}
\usepackage{color}
\usepackage{soul}
\usepackage[normalem]{ulem}
\usepackage{xcolor}
\definecolor{dblue} {RGB}{28,130,185}
\let\oldaddcontentsline\addcontentsline
\newcommand{\stoptocentries}{\renewcommand{\addcontentsline}[3]{}}
\newcommand{\starttocentries}{\let\addcontentsline\oldaddcontentsline}
\definecolor{nred}{RGB}{224,0,0}
\definecolor{nblue}{RGB}{28,130,185}
\definecolor{darkgreen}{rgb}{0,0.60,.2}
\definecolor{pgreen}{rgb}{0,1.0,0.}

\newcommand\redsout{\bgroup\markoverwith{\textcolor{brown}{\rule[0.6ex]{4pt}{1.0pt}}}\ULon}
\begin{document}
\title{Emergent dipole moment conservation and subdiffusion in tilted chains}

\author{S. Nandy}
\affiliation{Jo\v zef Stefan Institute, SI-1000 Ljubljana, Slovenia}
\author{J. Herbrych}
\affiliation{Institute of Theoretical Physics, Faculty of Fundamental Problems of Technology, Wroc\l aw University of Science and Technology, 50-370 Wroc\l aw, Poland}
\author{Z. Lenar\v{c}i\v{c}}
\affiliation{Jo\v zef Stefan Institute, SI-1000 Ljubljana, Slovenia}
\author{A. G\l\'odkowski}
\affiliation{Institute of Theoretical Physics, Faculty of Fundamental Problems of Technology, Wroc\l aw University of Science and Technology, 50-370 Wroc\l aw, Poland}
\author{P. Prelov\v{s}ek}
\affiliation{Jo\v zef Stefan Institute, SI-1000 Ljubljana, Slovenia}
\author{M. Mierzejewski}
\affiliation{Institute of Theoretical Physics, Faculty of Fundamental Problems of Technology, Wroc\l aw University of Science and Technology, 50-370 Wroc\l aw, Poland}

\date{\today}
\begin{abstract}
We study the transport dynamics of an interacting tilted (Stark) chain. We show that the crossover between diffusive and subdiffusive dynamics is governed by $F\sqrt{L}$, where $F$ is the strength of the field, and $L$ is the wave-length of the excitation. While the subdiffusive dynamics persist for large fields, the corresponding transport coefficient is exponentially suppressed with $F$ so that the finite-time dynamics appear almost frozen. We explain the crossover scale between the diffusive and subdiffusive transport by bounding the dynamics of the dipole moment for arbitrary initial state. We also prove its emergent conservation at infinite temperature.
Consequently, the studied chain is one of the simplest experimentally realizable models for which numerical data are consistent with the hydrodynamics of fractons.  
\end{abstract}
\maketitle
\stoptocentries
{\it Introduction.}  For generic closed many-body systems, the out-of-equilibrium dynamics results in a relaxation into a state of equilibrium specified by the conserved quantities. In such systems, the long-wavelength excitations associated with these quantities attenuate following the near-universal Fick's law of diffusion. For more than a decade, much attention has been devoted to systems that could violate these ubiquitous properties, displaying anomalous diffusion or even failing to thermalize entirely.  Tilted (Stark) quantum systems, subjected to a linear potential, offer a concrete and experimentally accessible platform to explore these extraordinary phenomena. Such systems can be realized in cold-atom experiments \cite{Guardado2020, Scherg2021, Kohlert2023}, and their properties have served as one of the main motivations for theoretical studies concerning Stark many-body localization (SMBL), Hilbert space fragmentation, and fracton hydrodynamics. 


SMBL emerged as a phenomenon that was expected to exhibit physics resembling the conventional many-body localization (MBL) in that single-particle Stark localization survives despite the presence of interactions between particles \cite{Schulz2019, Nieuwenburg2019,Taylor2020}.  Later on, SMBL was studied theoretically for various models \cite{Chanda2020,Bhakuni2020,Kshetrimayum2020,Zhang2021}  and experimentally also in a trapped-ion quantum simulator \cite{Morong2021}. Despite a similarity to  MBL, the nonergodicity of  SMBL systems has a distinct physical origin \cite{Scherg2021,Doggen2021}  and is expected to be transient,  at least in finite systems \cite{Zisling2022}. In the case of large tilt, the  Schrieffer-Wolff transformation allows one to derive (approximate) effective Hamiltonians, which strictly conserve the dipole moment \cite{Scherg2021,Bernevig2019}.  Combination of the particle-number and dipole conservations leads to extensive fragmentation of the Hilbert space  \cite{Rakovszky2020,Feldmeier2021,Moudgalya2022_1}  and to a breakdown of thermalization \cite{Khemani2020,Sala2020,Morningstar2020,Pai2020,Moudgalya2022}.

On the one hand,  the dipole moment is not strictly conserved in experimentally relevant tilted models.  Therefore, some studies indicate that the nonergodicity of tilted chains is only a prethermal phenomenon, which is eventually replaced by thermalization \cite{Khemani2020,Scherg2021,Kohlert2023}. On the other hand,  fluids in which charge and dipole moment are  conserved \footnote{Such systems obey a generalized continuity equation $\partial_t n + \partial^2_x \mathcal J=0$, where $n$ represents the charge density, and $\mathcal J$ denotes the local current, which can be interpreted as a flux of dipoles. Applying the hydrodynamic paradigm, we can write down an expression for $\mathcal J$ in terms of the charge density organized in a perturbative gradient expansion and solve the equation order by order in the derivatives. In fact, the leading order term is $\mathcal J = D\partial^2_x n$, which leads to the generalized Fick's law $\partial_t n + \partial^2_x (D \partial^2_x n)=0$.} exhibit unconventional transport properties, as described by the framework of fracton hydrodynamics \cite{Nandkishore2019,Grosvenor2022,PhysRevE.107.034142}. One of the distinctive features of dipole-conserving systems is the prediction of subdiffusive relaxation of charge-density modulation. A modulation with wavevector $q$ relaxes with the rate $\Gamma \propto q^4$ \cite{Gromov2020,Feldmeier2020,Burchards2022}, as observed experimentally in strongly tilted planar cold-atom lattices \cite{Guardado2020}.  


A one-dimensional (1D) tilted chain with short-range interactions is the simplest system in which these phenomena may possibly occur and which is closely related to the experimental setups. However, so far, it is not clear how to properly reconcile the presence of seemingly conflicting phenomena: the subdiffusive relaxation (with rate $\Gamma = D q^4$), originating from the conservation of the dipole moment, the transient absence of ergodicity originating from nearly fragmented Hilbert space, and the possible asymptotic thermalization originating from the fact that the dipole moment is not strictly conserved but is rather a time-dependent quantity. 

In this Letter, we study interacting spinless fermions on chains with  $L$ sites tilted by the electric field $F$.  We present compelling numerical evidence supporting the subdiffusive relaxation of the hydrodynamic density modes $A_q \propto \exp(-D q^z t)$.  Upon increasing $F$, for the smallest considered $q \sim 1/L$, the exponent changes from the diffusive value $z=2$ to the subdiffusive $z=4$ and, quite unexpectedly,  the crossover in $z$ is determined by the magnitude of  $F\sqrt{L}$. Moreover, the transport coefficient $D$ in the subdiffusive regime decreases exponentially with the field, which can be reconciled with previous observations of nonergodicity for strong $F$. The diffusion-to-subdiffusion crossover can be linked with the emergent conservation of the dipole moment that is shown to hold true also for nonequilibrium evolution starting from an arbitrary initial state. In the case of infinite temperature, one can prove a stronger bound on the time-dependence of the dipole moment. The latter bound holds for a broad class of models and also for multidimensional systems.


\begin{figure}[!tb]
\includegraphics[width=1.0\columnwidth]{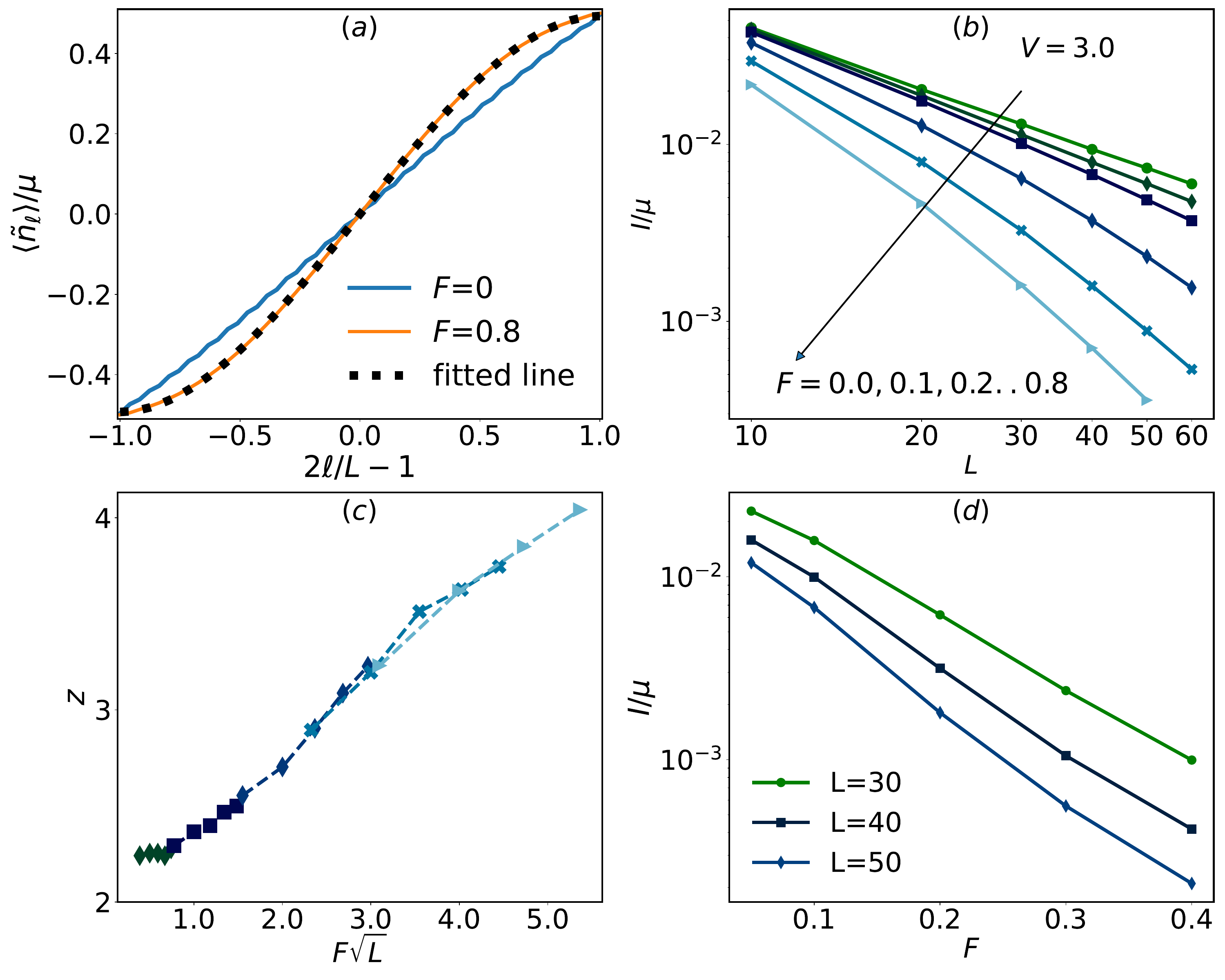}
\caption{Results for NESS in chains with boundary driving. (a) Normalized amplitude of the spatial profile of particle density for $L=50$ and two different tilts: $F=0$ (diffusive) and $F=0.8$ (subdiffusive). A cubic profile, $\langle \tilde{n}_l\rangle=a x+bx^3, x=2l/L-1$, consistent with the hydrodynamic equation corresponding to the $z=4$ case, fits the data well for $F=0.8$.  (b) NESS current $I/\mu$ vs. $L$ for different values of field $F$, where $F \in [0.0, 0.1, 0.2, 0.4, 0.6, 0.8]$. The arrow points to the direction of increasing $F$. (c) Exponent $z$ as a function of $F\sqrt{L}$. (d) NESS current $I/\mu$ reveals the exponential decay with $F$.}
\label{fig1}
\end{figure}

{\it Steady-state properties.}  We first study a Stark chain with $L$ sites and open boundary conditions (obc). It contains $L/2$ spinless fermions interacting via nearest-neighbor (nn) repulsion $V$, 
\begin{eqnarray}
H & = & H_0+F M \; , \nonumber \\
H_0 &= &\sum_l \left( c^{\dagger}_l  c_{l+1}+c^{\dagger}_{l+1}  c_{l} \right)  +V \sum_l  \tilde n_{l} \tilde n_{l+1}   +H' \;.  \label{ham}  
\end{eqnarray}
Here, $c^{\dagger}_l$ creates a particle at site $l$,  $n_{l}=c^{\dagger}_l c_l$, $\tilde{n}_{l} = n_{l}-1/2$, and $M =\sum_l (l-L/2) \tilde{n}_{l}$ is the dipole moment. We have added a term $H'$ that breaks the integrability of the model at $F=0$, allowing for normal diffusion at weak $F$, and its choice is optimized for the applied numerical method. In the main text, we discuss numerical results for $V=3$, whereas results for smaller $V$ are shown in the Supplementary Material, Ref.~\cite{supmat}.

First, we study an open chain, Eq.~\eqref{ham}, that is driven via boundary Lindblad operators with a small particle current injection rate $\mu$ \cite{Znidaric2011}. We employ the time-evolving block decimation (TEBD) technique for vectorized density matrices \cite{Verstraete2004,Zwolak2004} to solve the Lindblad master equation.  It allows us to establish the nonequilibrium steady state (NESS) for which we calculate the normalized particle current, $I/\mu$, and the spatial profile of particles, $\langle \tilde{n}_{l} \rangle$. When using the TEBD, it is convenient to stay within the nearest-neighbor interaction. Therefore, to break the integrability at $F=0$, we resort to a term having the form $H'= \tilde{V}\sum_{l}(-1)^{l}\tilde{n}_{l} \tilde{n}_{l+1}$. We set $\tilde{V}=0.4$ throughout the paper unless mentioned otherwise. All other details concerning numerical calculations are explained in Ref. \cite{supmat}.

Fig.~\ref{fig1}(a) shows the rescaled steady-state density profiles, $\langle \tilde n_l \rangle/\mu$, for two different field strengths $F$ with a system size of $L=50$. Diffusive systems follow the standard hydrodynamic equation, $\partial_t n-D \partial^2_x n=0$, characterized by a linear steady-state profile, $\partial^2_x n=0$. In Fig.~\ref{fig1}(a), we indeed see a linear profile at $F=0$, consistent with the exponent being $ z = 2$. However, for a large enough field $F=0.8$, the numerical results clearly reveal the nonlinearity of the profile. Such pronounced non-linear profile can be very accurately fitted to a  cubic function, $\langle \tilde{n}_l\rangle=a x+bx^3, x=2l/L-1$, as shown in Fig.~\ref{fig1}(a). Note that a cubic profile of the steady state is consistent with the generalized hydrodynamic equation for systems conserving the total charge and dipole moment \cite{Note1}, $\partial_t n+D \partial^4_x n=0$, from which one obtains $\partial^4_x n=0$ and $z=4$.


To study the crossover from diffusive ($z=2$) to subdiffusive ($z=4$) regimes in more detail, we have calculated $L$-dependence of normalized particle current $I/\mu$ shown in Fig.~\ref{fig1}(b). The dependence of the dynamical exponent $z$ on $L$ is manifested via the bending of the curves, particularly pronounced for the larger $F$. To extract the exponent $z$, related to the relaxation of the slowest modes on the system of size $L$, we fit the slope between two consecutive calculated points on the plot  $\log(I/\mu)$ vs. $\log(L)$ and locally assume scaling $I\sim L^{1-z}$. Fig.~\ref{fig1}(c) shows that extracted $z$ collapse on a single curve when plotted as a function of $F \sqrt{L}$. Later on, we explain the origin of this unexpected scaling. It indicates that macroscopic chains  ($L \to \infty$) with nonzero $F$ are always subdiffusive, whereas diffusive behavior can be observed only in finite systems or shorter wave-lengths. 
Another important observation that follows directly from panel (d) of Fig. \ref{fig1} is that the normalized current $I /\mu$ for a fixed $\mu$ decays exponentially with $F$. This immediately implies that for a very strong $F$, the studied systems would appear localized and nonergodic.


\begin{figure}[!tb]
\includegraphics[width=1.0\columnwidth]{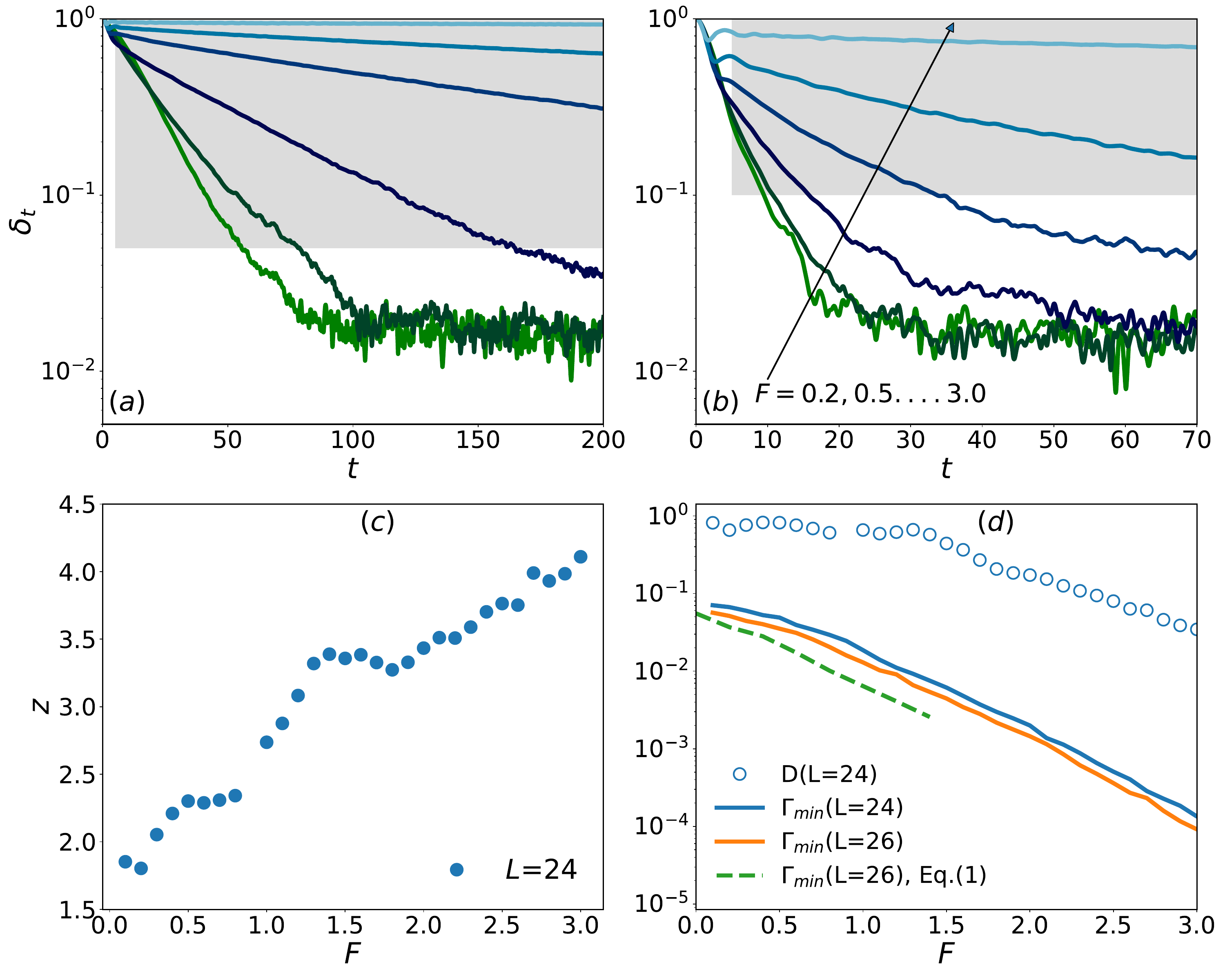}
\caption{Panel (a) and (b) depicts time evolution of normalized amplitudes, $\delta_t$,  of the spatially modulated profile for $L=24$ with the wave-vectors $q=2 \pi/L$ and $q=4\pi/L$, respectively. For both panels: $F \in [0.2, 0.5, 1.0, 1.5, 2.0, 3.0]$ and the arrow points to the direction of increasing $F$. Panels (c) and (d) respectively show the field $F$ dependence of the exponent $z$ and that of the transport coefficient $D$ alongwith the decay rate, $\Gamma_{min}= Dq^z$, for the smallest $q=2\pi/L$. Results in (c) and (d) are obtained from exponential fits to data in the shaded areas in (a) and (b). (d) Also shows the $\Gamma_{min}$ obtained for the Hamiltonian~(\ref{ham}) with obc (see the Supplementary Material \cite{supmat} for the details).}
\label{fig2}
\end{figure}

{\it Dynamics of the density modulations.} 
Next we confirm that conclusions formulated from the NESS are consistent also with the time-evolution of initial states with spatially modulated densities of particles, \mbox{$\delta n_l=  \langle \tilde n_l \rangle \propto \cos(q l) $} with \mbox{$q = 2 \pi m/L \ll 1$} and  $m=1,2$. The evolution is obtained via the Lanczos propagation method \cite{lantime1,lantime2}, which allows to study chains with up to  $L=26$.  In order to reduce boundary effects,  the time-evolution is carried out for a system equivalent to Eq.~(\ref{ham}), but with time-dependent flux and periodic boundary conditions
\begin{equation}
H_F(t) =\sum_l \left( e^{-iFt }c^{\dagger}_l  c_{l+1}+ e^{iFt }c^{\dagger}_{l+1}  c_l +V  n_{l} n_{l+1}  \right)+H' \label{ham_t} \; .
\end{equation}
Here, it is more convenient to choose another form of the integrability-breaking term that does not affect the translational symmetry, namely $H'=V' \sum_l n_{l} n_{l+2}$ where we take $V'=2$.

We start the time-evolution from a microcanonical thermal state obtained numerically \cite{long03,prelovsek11,herbrych22} for the Hamiltonian \mbox{$H_{t<0}= H_{F=0}(t) + \sum_l \cos(q l) n_{l},$} where the first term is defined in Eq.~(\ref{ham_t}). In order to be in the linear response regime and to obtain small but nonzero particle-density modulation, we use large but finite temperature  $kT=1/\beta=10$. Within the high-temperature expansion one estimates $\langle \tilde{n}_l \rangle  \simeq -(\beta/4) \cos(q l) \ll 1/2$. At time $t=0$, we quench the electric field $F$ and determine evolution under the Hamiltonian  (\ref{ham_t}) with $F \ne 0$. We calculate $\delta n(t)=\sqrt{\langle \langle \tilde{n}_l \rangle^2 \rangle_l}$, where  $\langle ...\rangle_l$ denotes averaging over all lattice sites. Technical details are discussed in the Supplementary Material, Ref. \cite{supmat}.

Results in Figs.~\ref{fig2}(a) and  \ref{fig2}(b) show time-dependence of the ratio \mbox{$\delta_t= \delta n(t)/\delta n(0)$}  at different fields.  It decays (up to an additive finite-size constant \cite{supmat}) exponentially with time, as expected for systems with either diffusive or subdiffusive transport.  We therefore use the ansatz  $\delta_t=\exp(- \Gamma_q t ) $ with  $\Gamma_q = D q^z$. Numerical results for $\Gamma_q$ at two smallest wave vectors $q=2 \pi/L, 4 \pi/L$  determine the exponent $z$ and the  transport coefficient $D$   which are shown in Figs.~\ref{fig2}(c) and  \ref{fig2}(d), respectively. Moreover, results for the decay rates $\Gamma_{min} $ for smallest $q=2 \pi/L$ in Fig.~\ref{fig2}(d) reveal exponential dependence on $F$, and they also agree with results obtained from the density correlation-function analysis within the original tilted model, Eq.~(\ref{ham}), as described in Ref.~\cite{supmat}.

Results obtained from the NESS properties \mbox{(Fig. \ref{fig1})} as well as from the dynamics of the spatial profiles  \mbox{(Fig. \ref{fig2})} support a consistent picture of transport in the studied system. The crossover from diffusive transport  ($z=2$) to subdiffusive ($z=4$) depends on the strength of the field as well as on the length-scale of the modulation (or equivalently on the related wavelength $q \propto 1/L$) and follows the scaling $z=z(F\sqrt{L})$. While the transport remains subdiffusive even for strong $F$, the corresponding transport coefficient, $D$, is strongly suppressed, i.e., the decrease of $D$ with increasing $F$ is at least exponential. Latter is in agreement with a Floquet interpretation of the problem, where $F$ plays the role of large frequency, causing transitions that are exponentially suppressed in the number of excitations needed to absorb such a large energy \cite{gunawardana22,mori16,abanin17},
however, one should note that we observe exponential supression already at intermediate $F$. It should also be reminded that the Floquet Hamiltonian, Eq.~\eqref{ham_t}, and the tilted model, Eq.~\eqref{ham} are in finite systems equivalent only up to boundary terms. Furthermore, the $D(F)$ dependence of the transport coefficient also partially resembles the disorder-dependence of the diffusion constant in disordered interacting chains \cite{barisic10,barisic16,steinigeweg16,prelovsek17,prelovsek21,herbrych22,prelovsek2023}. As a consequence, finite systems appear almost localized for very strong fields in agreement with the previous numerical studies that suggested Stark many-body localization and absence of thermalization in the effective
models with fragmented Hilbert spaces.


{\it Bound on the dynamics of the dipole moment.}  The dynamics of the dipole moment remains an important open issue, which determines whether transport in the studied systems is correctly captured by the fracton hydrodynamics. A simple bound on the maximal variation of the dipole moment during the time-evolution can be obtained from the identity,
\begin{equation}
\frac{{\rm d } \langle M \rangle_t  }{{\rm d} t}= \frac{1}{F} \frac{{\rm d } \langle H-H_0 \rangle_t  }{{\rm d} t}=-\frac{1}{F} \frac{{\rm d } \langle H_0 \rangle_t  }{{\rm d} t}, \label{eqmaindep}
\end{equation} 
where $H$ and $H_0$ are introduced in Eq.~(\ref{ham}) while \mbox{$ \langle M \rangle_t  =\langle \psi(0) | \exp(iHt) M \exp(-iHt) |\psi(0) \rangle $} and $|\psi(0) \rangle $ is the initial state.
 The change of $\langle H_0 \rangle_t$  is limited by a span of eigenvalues of $H_0$ leading to bound that is linear in $L$, i.e., \mbox{$| \langle H_0 \rangle_t-\langle H_0 \rangle_{t'} \rangle| < \alpha L$.} Here $\alpha$ is $F$- and $L$-independent constant determined by parameters of $H_0$.  Then one gets from  Eq.~(\ref{eqmaindep}) also a bound on  the variation of the dipole moment \mbox{$| \langle M \rangle_t -\langle M \rangle_{t'}| < \delta_M= \alpha L/F$}.  The latter bound should be compared to the width of the spectrum of the dipole moment operator, \mbox{$M |\psi_n \rangle =  d_n |\psi_n \rangle $}.  The density of its eigenvalues $\rho(d)=\frac{1}{Z} \sum_n \delta(d-d_n)$ is described by a Gaussian with the variance 
\begin{equation}
\sigma^2_M=  \frac{1}{Z} {\rm Tr}(M^2)=\sum_{l=-\frac{L}{2}+1}^{\frac{L}{2}} \frac{l^2}{4} \simeq \frac{L^3}{48}. \label{width}
\end{equation}  
Consequently,  one gets  
\begin{equation}
\frac{\delta_M}{\sigma_M} = \frac{4 \alpha \sqrt{3}}{F \sqrt{L}}\; . \label{bound1}
\end{equation}
For arbitrary nonzero $F$ and sufficiently large $L$, changes in the dipole moment become negligible and the fractonic dynamics sets in. The ratio in Eq.~(\ref{bound1}) also explains the unexpected scaling shown for the crossover from normal diffusive transport to subdiffusive dynamics shown in Fig.~\ref{fig1}(c).  
We stress that Eq.~(\ref{bound1}) holds for arbitrary initial $| \psi(0) \rangle$ and hence is applicable also for nonequilibrium dynamics. This equation originates from the exceptionally broad spectrum of the dipole moment. Namely, the width of the spectrum of typical, extensive operators (e.g., $H_0$ or $H'$) increases as $L^{1/2}$ in contrast to $L^{3/2}$ dependence of $\sigma_M$ in Eq. (\ref{width}).  

{\it Emergent conservation of the dipole moment at infinite temperature.} 
In the case of dynamics at \mbox{$T\to \infty$}, one can formulate a stronger bound on the variation of $M$. Below we prove that  the dipole moment is conserved in macroscopic systems, i.e, 
\begin{equation}
\lim_{L \to \infty} \frac{\langle  M(t) M \rangle}{\langle M M \rangle}=1, \quad \quad M(t)=e^{i Ht } M e^{-i Ht} \;. \label{target}
\end{equation}
Eq. (\ref{target}) will be shown to hold at any time $t$ for arbitrary nonzero $F$. From now on, we use the symbol  $\langle ... \rangle$ to denote $T \to \infty$ average,  $ \langle ... \rangle= (1/Z) {\rm Tr(...)}$. We recall that $||A||^2=\langle A A \rangle$ is the (squared)  Hilbert-Schmidt (HS) norm of a Hermitian operator $A$ while  $\langle A B \rangle$ is the HS  inner product of Hermitian $A$ and $B$. We note also that two terms entering Hamiltonian (\ref{ham}) are mutually orthogonal, $\langle H_0 M \rangle=0$. Due to this orthogonality, one obtains
\begin{eqnarray}
||H||^2 & = &|| H_0||^2  + F^2  || M ||^2 , \label{norm1} \\
\frac{|| H_0||}{|| M ||} & \propto & \frac{L^{1/2}}{L^{3/2}}, \label{norm2} 
\end{eqnarray}
where we used Eq.~(\ref{width}). As a central step, we split the dipole moment into two  mutually orthogonal parts, \mbox{$M=M^{\parallel}+M^{\perp}$}, defined via the projection
\begin{eqnarray}
M^{\parallel}&=& \frac{\langle M H \rangle}{\langle H H \rangle} H, \quad \quad M^{\perp}=M-M^{\parallel}, \label{proj1} \\
|| M  ||^2&=&|| M^{\parallel}||^2+|| M^{\perp} ||^2.\label{proj2}
\end{eqnarray}
so that $M^{\parallel}$ is conserved by construction. Using \mbox{Eqs. (\ref{norm1}),(\ref{proj1}) and (\ref{proj2})} we calculate the  HS norms of both components of the dipole moment 
\begin{eqnarray}
||  M^{\parallel} ||^2 &=&\frac{\langle MH \rangle^2}{|| H ||^2}=
||M||^2  \frac{F^2 || M||^2}{|| H_0||^2 +F^2|| M||^2} \nonumber \\
|| M^{\perp} ||^2 & = & ||M||^2  \frac{|| H_0||^2}{|| H_0||^2 +F^2|| M||^2} . \label{normp}
\end{eqnarray}
Finally, we obtain a  bound on the numerator in Eq.~(\ref{target})
\begin{eqnarray}
\langle [M^{\parallel} +M^{\perp}(t)] M \rangle  &\ge&  || M^{\parallel}  ||^2 -|\langle  M^{\perp}(t) M \rangle | \nonumber \\ 
&\ge& ||  M^{\parallel} ||^2-||M^{\perp}||\; ||M||, 
\end{eqnarray}
where we used the orthogonality $\langle M^{\parallel} M^{\perp} \rangle =0$ as well as the Cauchy–Schwarz (CS) inequality for \mbox{$|\langle  M^{\perp}(t) M \rangle|$}. The CS inequality also implies that the correlation function in Eq. (\ref{target}) is not larger than unity, hence
\begin{equation}
1\ge\frac{\langle M(t) M \rangle}{||M||^2} \ge 1- \frac{||M^{\perp}||}{||M||}- \frac{||M^{\perp}||^2}{||M||^2}.\label{bound2}
\end{equation} 
These inequalities together with Eqs.(\ref{norm2}) and  (\ref{normp}) prove Eq. (\ref{target}). The details of the model enter the above derivation only via Eqs.~(\ref{norm1}) and  (\ref{norm2}), which hold true for a broad class of Hamiltonians. Moreover, the same reasoning is applicable also for multidimensional systems with $L$ sites when \mbox{$|| H_0 || \propto L^{1/2}$} while   $||M|| \propto L_F L^{1/2}$, where $L_F$ is the size of the system along the field. Then the ratio in Eq.~(\ref{norm2}) vanishes when $L_F$ becomes infinite.

We note that the bound on the variation of the dipole moment for a chain at $T\to \infty$ (\ref{bound2}) is stronger than the one obtained in Eq. (\ref{bound1}). Therefore, the emergent conservation of the dipole moment at infinite temperature takes place for much weaker fields, $F \gg 1/L$ (see also Refs. \cite{Guardado2020, Zechmann_Knap_2022}), compared with dynamics starting from an arbitrary nonequilibrium state where one needs $F\gg1/\sqrt{L}$.

{\it Concluding remarks} We have studied tilted (Stark) chains with short-range interaction. Such a system has previously been studied as a prototype model for several phenomena ranging from the Stark many-body localization, subdiffusive relaxation of the density profiles, and the Hilbert space fragmentation found in the effective models. Our numerical studies provide a coherent and unifying picture that captures these phenomena as well as quantitative results for the dynamics of the tilted chains. We have found that the relaxation rate, $\Gamma_q = Dq^z$, of the density profile with the wave-vector $q\sim 1/L$  exhibits a crossover from a diffusive behavior, $z=2$, for small $F\sqrt{L}$ to the subdiffusive one, $z=4$ when $F \sqrt{L}$ is large. The subdiffusive transport coefficient $D$ is exponentially suppressed for strong fields $F$ so that the finite-time dynamics can hardly be distinguished from a nonergodic behavior. The exponential suppression of $D$ can be explained via analogy with driven Floquet systems as used in Eq.~\eqref{ham_t}. Nevertheless, we have not found any signature of strict nonergodicity expected in SMBL as well as in the effective models with strictly conserved dipole moment and finite-range interaction.
 
The unexpected scaling $z=z(F\sqrt{L})$ can be linked to the fraction of the spectrum of the dipole moment, $M$, that is accessible to the system during its evolution. It means that macroscopic chains (isolated in the bulk) subject to nonzero $F$ are subdiffusive (at small enough $q$) and that they evolve within a vanishing small part of the spectrum of $M$ independently of the initial state of the evolution. At the same time,  the dynamics of $M$ at $T\to \infty$ reveals the emergent conservation of the dipole moment [in the sense of Eq. (\ref{target})], formally shown for a broad class of models and also for multidimensional systems. Our results indicate that an interacting tilted chain can be considered one of the simplest systems that realize fractonic hydrodynamics; however, the corresponding transport coefficient is exponentially suppressed by strong fields.

\acknowledgements 
S.N. and Z.L. thank Marko Ljubotina, Michael Knap for illuminating discussions.
We acknowledge the support by the National Science Centre, Poland, via project 2020/37/B/ST3/00020 (M.M.), 2019/34/E/ST3/00405 (A.G.) and the support by the Slovenian Research Agency via the program P1-0044 (P.P), projects J1-2463, N1-0318 (Z.L and S.N.). Z.L and S.N. were supported also by QuantERA grants QuSiED and T-NiSQ, by MVZI, QuantERA II JTC 2021. The numerical calculations were carried out at the facilities of the Wroclaw Centre for Networking and Supercomputing, at the Spinon cluster of Jo\v zef Stefan Institute, Ljubljana, and supercomputer Vega at the Institute
of Information Science (IZUM) in Maribor, Slovenia. Our TEBD code was written using ITensors Library in Julia. \cite{ITensor}.
 

\newpage
\phantom{a}
\newpage
\setcounter{figure}{0}
\setcounter{equation}{0}

\renewcommand{\thetable}{S\arabic{table}}
\renewcommand{\thefigure}{S\arabic{figure}}
\renewcommand{\theequation}{S\arabic{equation}}
\renewcommand{\thepage}{S\arabic{page}}

\renewcommand{\thesection}{S\arabic{section}}

\onecolumngrid

\begin{center}
{\large \bf Supplemental Material:\\
Emergent dipole moment conservation and subdiffusion in tilted chains}\\

\vspace{0.3cm}

\setcounter{page}{1}

\ S. Nandy$^{1}$, J. Herbrych$^{2}$, Z. Lenar\v{c}i\v{c}$^{1}$, 
A. G\l\'odkowski$^{2}$, P. Prelov\v{s}ek$^{1}$, M. Mierzejewski$^{2}$
\\
\ \\
$^1${\it J. Stefan Institute, SI-1000 Ljubljana, Slovenia} \\
$^2${\it Institute of Theoretical Physics, Faculty of Fundamental Problems of Technology, \\ Wroc\l aw University of Science and Technology, 50-370 Wroc\l aw, Poland}\\
\end{center}

\vspace{0.6cm}
In the Supplemental Material we provide technical details for the (i) TEBD studies of an open system with Hamiltonian described by the Eq.~(\ref{ham}), (ii) the Lanczos-time propagation of density profiles in chains described by the time-dependent Hamiltonian in Eq.~(\ref{ham_t}), and (iii) density-correlation analysis of tilted systems with open boundary conditions, Eq.~(\ref{ham}).\\
\vspace{0.3cm}

\twocolumngrid

\label{pagesupp}

\section{TEBD studies of the boundary-driven system} 
For convenience of implementation, we resort to the spin version of the model described by the Hamiltonian in Eq.~\eqref{ham} The basic idea, in the spin language, is to drive a spin current across the system via boundary  Lindblad operators with a small spin bias $\mu$. To this end, the boundary Lindblads employed are of the form $L_1=\sqrt{1+\mu}S_{1}^{-}, L_2=\sqrt{1-\mu}S_{1}^{+}, L_3=\sqrt{1-\mu}S_{L}^{-}, L_4=\sqrt{1+\mu}S_{L}^{+}$. The master equation governing the evolution of the system's density matrix is given by
\begin{eqnarray}
\partial_{t}\rho=-i[H,\rho] + \hat{\mathcal{D}}\rho
\label{master_equ}
\end{eqnarray}
where $H$ denotes the Hamiltonian and $\hat{\mathcal{D}}$ stands for the dissipator expression in terms of the Lindblad operators as $\hat{\mathcal{D}}\rho=\sum_{k}L_{k}\rho L_{k}^{\dagger}-\frac{1}{2}\{L_{k}^{\dagger}L_{k}, \rho\}$. To evolve the density matrix towards the steady state $\rho_{ss}$, we use the time-evolving block decimation for vectorized density matrices. In particular, we use the fourth-order TEBD with a time step $dt = 0.2$, bond dimension $\chi \sim 140$, and bias $\mu \sim 0.01$.

\begin{figure}[!tb]
\includegraphics[width=1.0\columnwidth]{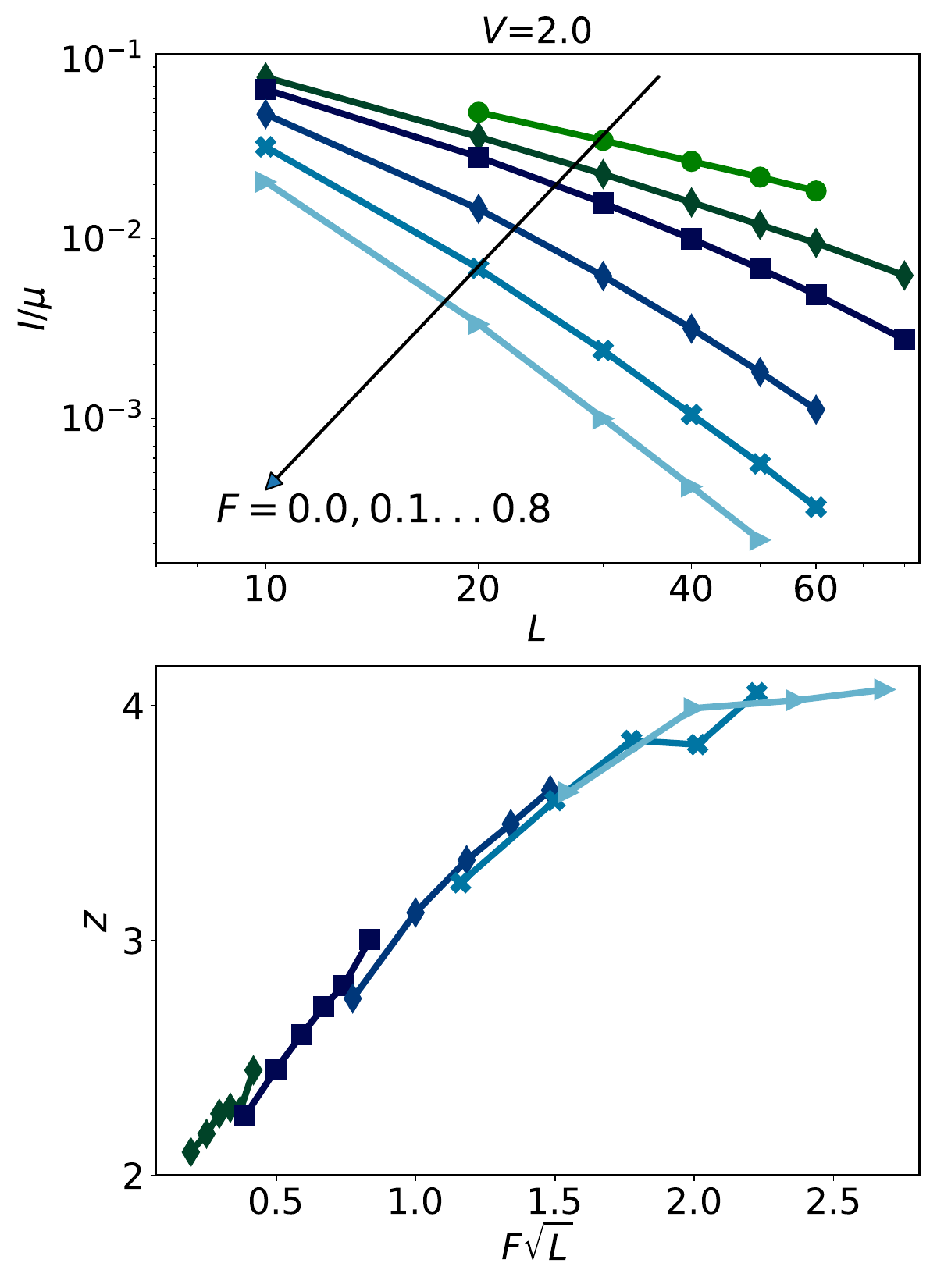}
\caption{The same as in Fig. \ref{fig1} (panel (b) and (c)) in the main text but for smaller $V=2$. As in Fig. \ref{fig1}, here also $F \in [0.0, 0.1, 0.2, 0.4, 0.6, 0.8]$, and the arrow in the upper panel points to the direction of increasing $F$.}
\label{sup1}
\end{figure}

Figure \ref{fig1} in the main text shows results obtained from TEBD studies for the chain described by the Hamiltonian (\ref{ham}) with $V=3$. In Fig. \ref{sup1}, we present similar data, but for weaker interactions $V=2$, leading to the same conclusions as presented in the main text.

\section{Relaxation of the density profiles for time-dependent Hamiltonian}
We start the time-evolution from a state obtained from the microcanonical Lanczos method (MCLM) \cite{long03,prelovsek11,herbrych22}. The method employs the Lanczos algorithm for solving the eigenproblem of the operator $(H_{\rm ini}-E_{\rm target})^2$. Here, the initial Hamiltonian reads \mbox{$H_{\rm ini}= H_{F=0}(t) + \sum_l \cos(q l) n_{l},$} and $H_{F=0}(t)$ is defined in Eq.~(\ref{ham_t}) in the main text. Since we are interested in the high-temperature regime, the energy of the microcanonical window, $E_{\rm target}$, is obtained from the high-temperature expansion, \mbox{$E_{\rm target}=\langle H_{\rm ini} \rangle_{\infty}-\beta \left[ \langle H^2_{\rm ini} \rangle_{\infty}-\langle H_{\rm ini} \rangle^2_{\infty} \right]$}. For simplicity, the infinite-temperature averages \mbox{$\langle ... \rangle_{\infty}$} are estimated from the grand canonical ensemble, and we set $\beta=0.1$. Using $N_s=5000$ Lanczos steps, we obtain an initial state $|\psi(0) \rangle$ that is a superposition of eigenstates of $H_{\rm ini}$ with energies centered at $E_{\rm tagret}$ and with a small energy spread $\Delta^2 E \propto L/N_s$.

The distribution of particles in the initial state is spatially modulated, $\langle \psi(0) | \tilde n_l |\psi(0) \rangle \simeq \ -(\beta/4) \cos(ql)$. After obtaining the initial state, we quench the field $F\ne 0$ and propagate this state under the Hamiltonian Eq.~(\ref{ham_t}) from the main text.Applying the Lanczos propagation method \cite{lantime1,lantime2} we numerically solve the time-dependent Schr\"odinger equation in small time-windows $\Delta t=0.01$ using 20 Lanczos steps for each time interval \mbox{$|\psi(t) \rangle \to |\psi(t+\Delta t ) \rangle $} and calculate the time-dependence of the profile \mbox{$\langle \tilde{n}_l \rangle =\langle \psi(t) | \tilde n_l |\psi(t) \rangle$}. Finally, we obtain the amplitude of the modulations \mbox{$\delta n(t)=\sqrt{\langle \langle \tilde{n}_l \rangle^2 \rangle_l}$}, where  $\langle ...\rangle_l$ denotes averaging over all lattice sites. Figs. \ref{fig2}(a) and  \ref{fig2}(b) in the main text, show the ratio \mbox{$\delta_t= \delta n(t)/\delta n(0)$} for $L=24$ and smallest wave-vectors $q_1=2\pi/L$ and $q_2=4\pi/L$, respectively.

\begin{figure}[!tb]
\includegraphics[width=1.0\columnwidth]{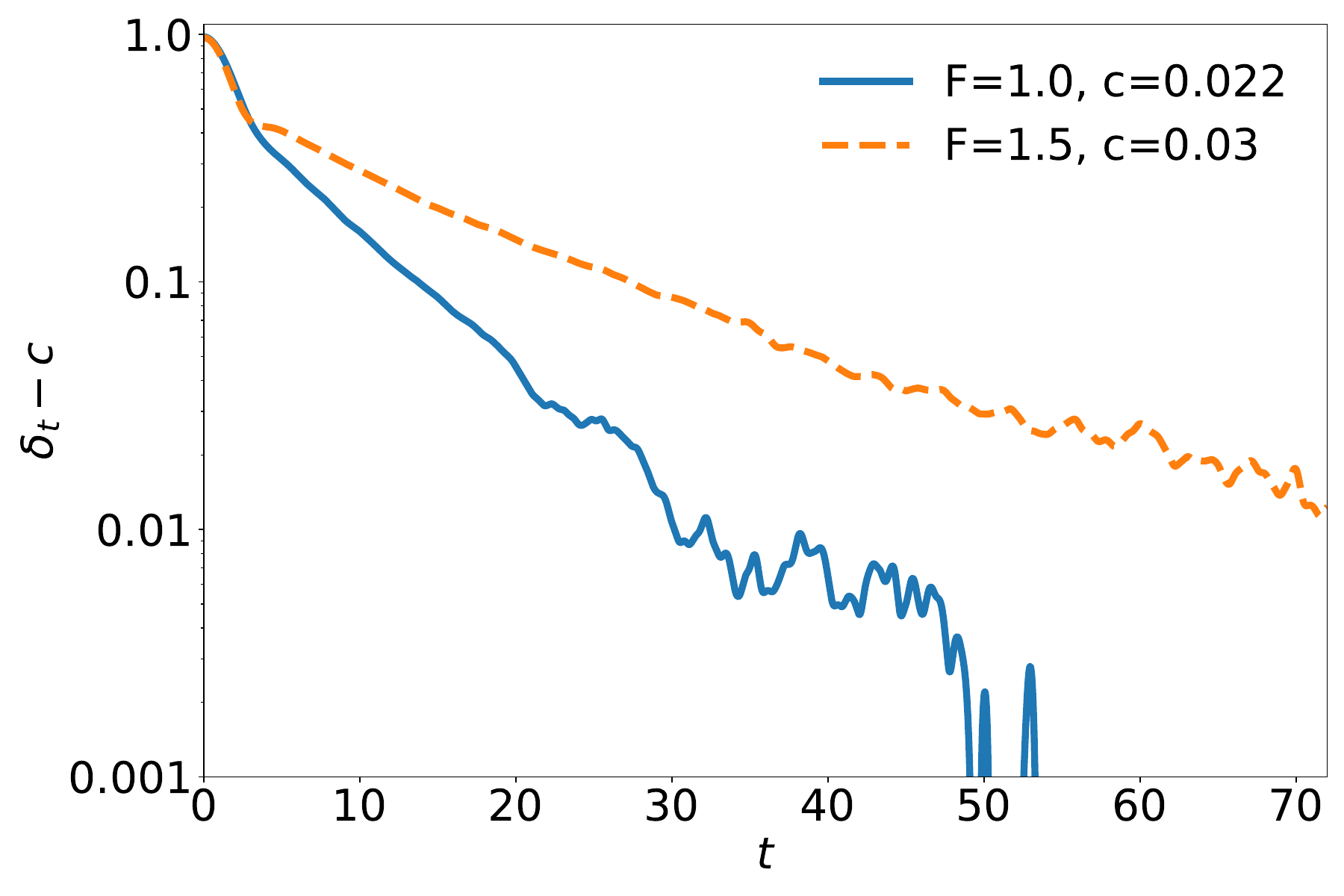}
\caption{Selected data from Fig. \ref{fig2}(b) in the main text shifted vertically by a constant $c$ indicated in the legend.}
\label{sup2}
\end{figure}

It is rather obvious that $\delta_t$ cannot strictly vanish in a finite system. Instead, one expects, 
$\delta_t=\exp(-\Gamma_q t)+c$, where $c$ is a constant, and we use the ansatz for the decay rate  $\Gamma_q=D q^z$. In the case of the profile with $q=q_1$, the constant $c$ is small enough and hardly affects the exponential decay of $\delta_t$; see Fig.~\ref{fig2}(a) in the main text. However, the offset ($c$) is larger for $q_2$, and some results in Fig. \ref{fig2}(b) in the main text might appear non-exponential. Therefore in Fig.~\ref{sup2} we show that $\delta_t-c$ is an exponential function for tuned, small $c$. 

In order to obtain the relaxation rate, $\Gamma_q$, one may either use a fitting function that includes the vertical offset, $c$, or one may restrict the range of fitted data to $\delta_t \gg c$. In order not to increase the number of fitting parameters, we choose the latter possibility and fit results for $t>5$ with $\delta_t>0.05$ and $\delta_t>0.1$ for $q=q_1$ and $q=q_2$, respectively. The fitted range of data is marked as a shaded area in Figs.\ref{fig2}(a) and \ref{fig2}(b) in the main text.  As a consistency check, we have also carried out fitting with the offset (not shown). The qualitative results remain unchanged. Upon increasing the tilt,  the exponent $z$ changes from $z=2$ to $z=4$ while the relaxation rate $\Gamma_q$, as well as the transport coefficient $D$ decreases exponentially for large $F$. 

\section{Numerical studies of a system with open boundary conditions}
In order to confirm the exponential decrease of the relaxation rate $\Gamma_q$ we have also carried out numerical studies of a tilted chain with open boundary conditions (obc), Eq.~\eqref{ham} in the main text, with the same $H'$ as for the time-dependent flux and same parameters $V=3, V'=2$. Results are shown in Fig.~\ref{fig2}(d) in the main text. In this case, the calculated quantity is the $T \to \infty$ dynamical structure factor $S(q,\omega)$, i.e., the density correlation function $\langle n_q(t) n_q(0) \rangle_\omega$ for $n_q = (1/\sqrt{L}$) $\sum_l \cos(q(l-L/2)) n_l$, again for smallest $q=2\pi/L$ and $q=4\pi/L$ (not shown). Results for the whole spectrum of $S(q,\omega)$ are obtained by employing MCLM using a large number of Lanczos steps (up to $N_L =8\times10^4$ for $L=26$) in order to reach high-frequency resolution (typically $\delta \omega < 10^{-3}$). MCLM at $T \to \infty$ in principle requires the averaging over the whole spread of the initial $E_{target}$, which was here due to substantial fluctuations somewhat restricted. To extract relaxation rates $\Gamma_q$ we use the low-$\omega$ representation 
\begin{equation}
S(q,\omega \sim 0)= \frac{-\chi^0_q}{\omega + i \Gamma_q}, \quad \chi^0_q= \frac{2}{\pi} \int  {\mathrm Im} S(q,\omega) d\omega
\end{equation}
where $\chi^0_q$ is static susceptibility, so that $\Gamma_q = \chi^0_q/ {\mathrm Im} S(q,\omega = 0)$.

Such an approach should yield the same result as the analysis via the time-dependent flux following Eq.~\eqref{ham_t}. Indeed, the comparison presented in Fig.~\ref{fig2}(d) in the main text confirms that. Still, in particular at the lowest $\omega \sim 0$ (or equivalently regarding long times $t \gg 1$), the results can be affected by obc, which restricts the reliable analysis to $\Gamma_q > 2.10^{-3}$, presented in Fig. \ref{fig2}(d).

\end{document}